# A Modest View of the Black Hole Information Paradox

Stephen Boughn*

Departments of Physics and Astronomy, Haverford College, Haverford, PA 19041

## Abstract

Thirty years ago, John Preskill [1] concluded "that the information loss paradox may well presage a revolution in fundamental physics" and mused that "Conceivably, the puzzle of black hole evaporation portends a scientific revolution as sweeping as that that led to the formulation of quantum theory in the early 20th century." Many still agree with this assessment. On the other hand, it seems to me the "paradox" has little to do with the physical world but rather, at best, simply points out the possible inconsistency of two, already disparate, theories (mathematical models) of nature, general relativity and quantum mechanics, with virtually no conceivable observational consequences. The information paradox hinges on the concepts of a pure quantum state, the unitarity of quantum mechanics, and Hawking's semi-classical calculation of black hole evaporation. I used the qualifier "at best" above because, for me, the concept of a quantum state is far more restrictive than required by the paradox while unitarity is not a property of nature but rather of a mathematical model and is certainly already violated by the process of making a measurement. Furthermore, the semi-classical calculation of Hawking is surely of limited applicability.

**Disclaimer:** I am an experimental physicist (now retired) and am ill-equipped to delve into the theoretical details of the black hole information paradox. At heart, I'm an empiricist and pragmatic to a fault. While I marvel at and have great respect for the wonderful mathematical models created by theorists, in the absence of observations my interest in them quickly wanes. In the last paragraph of Preskill's essay quoted above, he likewise expresses a "devout wish…that experiment can guide us"; however, in its absence, he muses, "…it is not so unrealistic to hope to make real progress via pure thought." As you might imagine, this is not my sentiment for how physics evolves. With this in mind, the purpose of my present essay is not to shed light on the information paradox but rather to explain my lack of interest in it.

## Introduction

The following is a thumbnail sketch of the black hole information paradox that I

---

* sboughn@haverford.edu

have gleaned from Preskill's essay [1], a 1997 *Scientific American* article by Lenny Susskind [2], and from various articles in popular journals (and, of course, the Wikipedia entry). I also "read" Hawking's original 1976 *Physical Review D* paper [3] but confess that I was unable to follow much of that paper.

The Schwarzschild solution of Einstein's general relativistic field equations is a spherically symmetry solution characterized by a physical singularity at the origin, a horizon at a finite radius (the Schwarzschild radius) and an approximately Newtonian gravitational field at larger radii. The horizon is the surface inside which no signal can propagate to the outside, hence, the name "black hole". The only physical property represented by this steady state solution is its mass, which is determined by the mass and energy that have collapsed to form the black hole. Black holes can also possess electrical charge and angular momentum, for which analogous solutions exists. The "no hair" theorem of general relativity proves that, after the steady state has been achieved, the only physical properties of black holes are mass, electrical charge, and angular momentum. So far so good. The singularity at the center of the black hole is certainly of concern but the consensus has always been it will be ultimately resolved by quantum mechanics.

In 1974 Hawking [4] demonstrated that a black hole posses an effective temperature and radiates thermal energy from its "surface" (near its horizon) thereby decreasing its mass, a process dubbed "black hole evaporation". Although there is, as yet, no viable quantum theory of gravity, Hawking's semiclassical derivation seems well founded and is accepted by most experts in the field. The evaporation rate is inversely proportional to the square of the black hole mass and is miniscule for typical nearby black holes with masses on the order of a few solar masses. The evaporation time for a 3 solar mass black hole is on the order of $10^{68}$ years, vastly larger than the lifetime of the universe, which renders black hole evaporation completely out of the realm of observational confirmation.

Even so, Hawking and many others since then have considered a gedanken experiment that points to an egregious violation of the "laws of physics". Suppose a black hole is formed from a collection of quantum particles in a definite (pure) quantum mechanical state[1]. Pure quantum mechanical states are characterized by maximum information, i.e., zero entropy. Such a black hole will "eventually" evaporate and therefore completely disappear with only the thermal Hawking radiation remaining. Therein lies the problem. The initial zero entropy (maximum information) pure quantum state seems to evolve into the maximum entropy (zero information) mixed quantum state of thermal Hawking radiation. On the other hand, the laws of quantum mechanics are unitary (time reversible) and, therefore, information preserving. Something must be wrong. Is it general relativity, which already has singularity problems, is it quantum theory, or are both theories in need of revision? This is the paradox that Preskill maintains presages a revolution in fundamental physics.

**Quantum States and Unitarity**

It may seem innocuous to contemplate forming a black hole from a unique, pure quantum state but one needs to look carefully at the meaning of such a declaration. Because my view of quantum states differs from that of many (most?) physicists, I will describe it in some detail. I subscribe to what is essentially Bohr's pragmatic view as espoused in Henry Stapp's 1972 paper, "The Copenhagen Interpretation" [5], which I read as a graduate student. In Stapp's practical account of quantum theory, a system to be measured is first prepared according to a set of specifications, $A$, which are then transcribed into a quantum state, a wave function $\Psi_A(x)$, where $x$ are the degrees of freedom of the system. The specifications $A$ are "couched in a language that is meaningful to an engineer or laboratory technician", i.e., not in the language of quantum (or even classical) formalism. Likewise, $B$ are a set of specifications of the subsequent measurement and its possible results. These are transcribed into another wave function $\Psi_B(y)$, where $y$ are the degrees of freedom of the measured system. How are the mappings of $A$ and $B$ to $\Psi_A(x)$ and $\Psi_B(y)$ effected? According to Stapp,

> …no one has yet made a qualitatively accurate theoretical description
> of a measuring device. Thus what experimentalists do, in practice,
> is to *calibrate* their devices…[then] with plausible assumptions…it
> is possible to build up a catalog of correspondences between what experimentalists do and see, and the wave functions of the prepared and measured systems. It is this body of accumulated empirical knowledge that bridges the gap between the operational specifications $A$ and $B$ and their mathematical images $\Psi_A$ and $\Psi_B$. Next a transition function $U(x; y)$ is constructed in accordance with certain theoretical rules…the 'transition amplitude' $\langle A|B \rangle \equiv \int \Psi_A(x) \, U(x; y) \Psi_B^* \, dxdy$ is computed. The predicted probability that a measurement performed in the manner specified by $B$ will yield a result specified by $B$, if the preparation is performed in the manner specified by $A$, is given by $P(A, B) = |\langle A|B \rangle|^2$.

---

[1] In nonrelativistic quantum mechanics, a pure state is characterized by a wave function that satisfies the Schrödinger equation.

What's my point here? From the pragmatic point of view, it is that the quantum mechanical wave function is a theoretical construct that we invented to deal with our observations of physical phenomena. Stapp's (and Bohr's) pragmatic account of wave functions is intimately tied to state preparation and measurement, both of which are described in terms of operational specifications that lie wholly outside the formalism of quantum mechanics. Prior to the preparation of a system the wave function is not even defined and after it has been measured the wave function ceases to have a referent.

This interpretation is to be contrasted with an ontological view of a quantum state. According to this realist perspective, all systems have unique quantum states independent of a physicist's knowledge of the state or intention to observe the system. The physicist's job is then to determine the state of a particular system. I suppose that for all practical purposes it doesn't matter whether one subscribes to Bohr's pragmatic interpretation or the realist's ontological interpretation of quantum mechanics. However, there are a handful of questions where it does matter. These include quantum state reduction (wave function collapse), the quantum measurement problem, and quantum nonlocality (Bell's theorem). Quantum realists are still struggling to find answers to these conundrums while quantum pragmatists claim they are meaningless questions. I'll argue in the next section that the quantum information paradox should be added to this list.

Because of the linearity of quantum mechanics, a superposition of two or more pure quantum states is also a pure quantum state. Another useful construct is a mixed quantum state, which is also alluded to in the information paradox. A mixed state is a statistical combination of pure quantum states. It cannot be represented by a pure quantum state, i.e., a wave function or vector in Hilbert space, but the density matrix of a mixed state is well-defined.  Mixed states are usually invoked in one of two situations: 1) when an observer has limited knowledge of a system and is only able to specify a statistical distribution of possible pure states; and 2) when the system is entangled with one or more remote quantum systems.  In the latter case, if one marginalizes over all possible states of the remote systems, the result is a well-defined mixed density matrix.

A quintessential entangled state consists of two spin ½ particles in a singlet (zero angular momentum) state.  According to quantum mechanics, their combined wave function (quantum state) is given by

$$\Psi(1,2) = \frac{1}{\sqrt{2}}\{|1,\uparrow\rangle|2,\downarrow\rangle - |1,\uparrow\rangle|2,\downarrow\rangle\} \qquad (1)$$

where $\uparrow$ and $\downarrow$ indicate the up and down components (in any direction) of the spins of particles 1 and 2.  This entangled state can equivalently be represented by a density matrix, $\rho(1,2) = |\Psi(1,2)\rangle\langle\Psi(1,2)|$.  Now marginalize over the possible states of particle 2 by taking the partial trace of the density matrix over a basis of system 2, i.e.,

$$\rho(1) = \langle 2,\uparrow| \, \rho(1,2)|2,\uparrow\rangle + \langle 2,\downarrow| \, \rho(1,2)|2,\downarrow\rangle. \qquad (2)$$

The result is $\rho(1) = \frac{1}{2}|1,\uparrow\rangle\langle 1,\uparrow| + \frac{1}{2}|1,\downarrow\rangle\langle 1,\downarrow|$, the density matrix of a mixed state of an unpolarized spin ½ particle and precisely the same as the state for which the system preparation can only be statistically specified, in this case a 50% probability that the particle is spin up and a 50% probability that the particle is spin down.

The unitarity of quantum mechanics implies time reversibility; a given pure quantum state evolves from a unique pure quantum state.  It is in this sense that unitary quantum mechanics is *information preserving*. On the other hand, no such inference is possible with regard to mixed quantum states; it is not possible to identify a quantum state from which a given mixed state has evolved.  Perhaps the most egregious breach of unitarity in quantum mechanics arises during the process of measurement.  For example, when the precise location of a particle is measured, there is no way to determine its prior quantum state.  Now, on to my quandary about the black hole information paradox.

**Interrogating the Information Paradox**

As an experimental physicist, I am naturally interested in how theoretical predictions can be observationally corroborated. With regard to the black hole information paradox this seems to be beyond the pale. After all, the smallest known black holes have evaporation times on the order of $10^{68}$ years, far beyond the lifetime of the universe. If there should happen to exist much smaller, primordial black holes, one might be able to observe their emitted Hawking radiation. Such black holes must have masses close to 5 x $10^{11}$ kg. Black holes much smaller than this would have evaporated long ago and black holes much larger than this will evaporate far in the future. It's not clear how primordial black holes would have formed but it would surely have been much earlier than the epoch of nucleosynthesis, a small fraction of a second after the big bang. Just how one might identify the quantum state of the constituents that formed them is not clear. In any case, in the last 50 years, there has been no astrophysical confirmation of such a system and the detailed observations required for the information paradox seems, to me, to be dubious at best.

On the other hand, the information paradox is posed in the context of a gedanken experiment and we know gedanken experiments are often extremely useful in elucidating physical phenomena. A prime example is Heisenberg's gamma ray microscope with which he derived his uncertainty principle. However, that gedanken experiment was closely associated with observable phenomena, photon interference and Compton scattering. The information paradox gedanken experiment has no such direct connection with observable phenomena. Putting this complaint aside, one is still left with the quandary of how one might specify the preparation of $\gtrsim 10^{58}$ particles that form a typical black hole. Of course, this requirement stems from my pragmatic view of a quantum state. The realist might claim that all these particles are automatically in a pure quantum state with no mention of state preparation.

Taking the latter view of quantum states, the question arises as to how this pure quantum state can possibly evolve into the mixed state of thermal Hawking radiation. This would contradict the unitarity of quantum mechanics. My question is: if the system of ~$10^{58}$ particles that form a black hole is automatically in a pure quantum state, why can't we also claim that the subsequent system of thermal Hawking radiation is in a pure quantum state and that the only reason it appears to be a mixed is because we are ignorant of the correlations between the radiation quanta? Hawking's analysis was semiclassical and it would naturally ignore such correlations. In addition, the notion that unitarity is a universal property of nature is suspect. Unitarity is a property of our theories (mathematical models) and shouldn't be accepted carte blanche as a property of nature herself. Of course, the realist might argue this is warranted because there are, indeed, *laws of nature* and our theories are reasonable approximations to those laws, in which case there is mounting evidence that the laws of nature are unitary. If this is the case, one might wonder why some processes, e.g., measurements, seem to violate unitarity and suspect that black hole processes might do the same.

Perhaps the information paradox doesn't imply that our understanding of nature is flawed but rather simply points to an inconsistency of our current mathematical models of nature. While this is a legitimate motivation for seeking a resolution, that it would "presage a revolution in fundamental physics" seems presumptuous to say the least. Even viewed in this light, I have my doubts about the efficacy of the analysis. Wouldn't a demonstration of the inconsistency of the model require a detailed calculation of the process in question. In the present case, this would require a specific Hamilton to evolve the in-falling pure quantum state until all the energy emerges via Hawking radiation. Of course we know of no such Hamiltonian. If we did and if it preserves unitarity, then presumably the outgoing state would conserve information, which Hawking radiation apparently fails to achieve. But this seems to be a rather strange argument. It assumes that some future model, which we don't yet posses, satisfies unitarity and if we could evolve the entire $10^{58}$ particle pure quantum state, which I'm quite sure we will never be able to do, then the results couldn't possibly consist of thermal Hawking radiation. And there still has been no mention of the physical singularity at the center of the black hole.

These are reasons why I personally don't take the information paradox seriously, but to be fair the theorists who have tackled the paradox have much more on their minds. Invariably they are interested in creating a unified quantum theory of all the fundamental forces in nature, a theory of everything, if you will. I'll make a few comments on this goal in my *Final Remarks* below.

**Final Remarks**

Am I suggesting that all the physicists who work on the information paradox are wasting their time? Certainly not! I've written two other essays [6,7] decrying analogous endeavors: Bell's inequality/quantum nonlocality; and Hugh Everett's many worlds interpretation of quantum mechanics. Although I've argued that these two efforts were ill-founded, many physicists have credited them with inspiration for pursuing new avenues of research. In the former case, many have attributed John Bell's 1964 paper with initiating both experimental and theoretical advances in quantum entanglement and thereby to progress in the new fields of quantum information and quantum computing (although, I'm less sanguine about the latter claim). In the many worlds case, while Everett was motivated by what I consider to be non-existent problems in quantum measurement, many people credit him with inspiring them to pursue research in new areas of physics, including decoherence theory, quantum information, and the application of quantum mechanics to cosmology [8]. I can't argue with these physicists or even pretend to know what motivates their endeavors, theoretical or experimental. A discussion of the sources of scientific creativity is certainly beyond my poor powers and I won't even attempt to address the topic.

There have been similar claims regarding the pursuit of resolutions to the black hole information paradox, i.e., that these pursuits have led to important contributions in the quest to create a quantum theory of gravity and thereby unify all the forces of nature. According to Suskind, "The information paradox, which appears to be well on its way to being resolved, has played an extraordinary role in this ongoing revolution [a quantum theory of gravity] in physics." For him, String Theory will likely provide the resolution. However, these goals were already being vigorously pursued and the information paradox simply provided a relevant focal point for some of the endeavors. Since there is currently no resolution of any of these issues, it's not clear how useful the information paradox will prove to be. In fact, the goal of quantizing gravity is, so far, only a mathematical pursuit with no experimental evidence in sight. Therefore, it's not surprising that these efforts are not particularly interesting to an empiricist like me. However, remember that the purpose of my essay is not a critical review of the progress in resolving the black hole information paradox but rather is to explain why it is of little interest to me.